\documentclass[twocolumn, 12pt]{article}%
\usepackage{amsmath,latexsym}
\usepackage{graphicx}
\usepackage{amsmath}
\usepackage{amsfonts}
\usepackage{amssymb}%
\begin{document}

\title{{Revisiting galactic rotation curves given a
   noncommutative-geometry background}}
   \author{
  Peter K.F. Kuhfittig and Vance D. Gladney* \\
  \footnote{kuhfitti@msoe.edu}
 \small Department of Mathematics, Milwaukee School of
Engineering,\\
\small Milwaukee, Wisconsin 53202-3109, USA}

\date{}
 \maketitle

\begin{abstract}\noindent
It was shown earlier by Rahaman et al. that a
noncommutative-geometry background can account for
galactic rotation curves without the need for dark
matter.  The smearing effect that characterizes
noncommutative geometry is described by means of
a Gaussian distribution intended to replace the
Dirac delta function.  The purpose of this paper
is two-fold: (1) to account for the galactic
rotation curves in a more transparent and
intuitively more appealing way by replacing the
Gaussian function by the simpler Lorentzian
distribution proposed by Nozari and Mehdipour
and (2) to show that the smearing effect is
both a necessary and sufficient condition for
meeting the stability criterion.\\

\noindent
PAC numbers: 04.20.Jb, 04.50.Kd
\\
\\
\textbf{Keywords:} Galactic Rotation Curves;
Noncommutative Geometry
\\
\\
\end{abstract}

\section{Introduction}\label{E:introduction}

That noncommutative geometry can account for
galactic rotation curves without the need for
dark matter has already been shown in Ref.
\cite{fR12}.  The effect in question is a small
effect, not only difficult to determine but
also difficult to present in an intuitively
appealing way.  This paper uses a slightly
different approach that may provide a clearer
picture.  This approach is introduced in
Sec. \ref{S:new} followed by the analysis in
Sec. \ref{S:solution}.  It is shown in Sec.
\ref{S:necessary} that the
noncommutative-geometry background is needed
for stability.

\section{Noncommutative geometry and galactic
rotation curves}\label{S:new}

An important outcome of string theory is the
realization that coordinates may become
noncommuting operators on a $D$-brane \cite
{eW96, SW99}.  The commutator is
$[\textbf{x}^{\mu},\textbf{x}^{\nu}]=
i\theta^{\mu\nu}$, where $\theta^{\mu\nu}$
is an antisymmetric matrix.  As discussed in
Refs. \cite{SS1, SS2}, noncommutativity
replaces point-like structures by smeared
objects.  The smearing effect is accomplished
by using a Gaussian distribution of minimal
length $\sqrt{\theta}$ instead of the Dirac
delta function \cite{NSS06, pK13}.  A simpler
but equally effective way is to assume that
the energy density of the static and
spherically symmetric and particle-like
gravitational source has the form \cite
{NM08, pKnew}
\begin{equation}\label{E:rho}
  \rho(r)=\frac{M\sqrt{\theta}}
     {\pi^2(r^2+\theta)^2}.
\end{equation}
Here the mass $M$ of the particle is diffused
throughout the region of linear dimension
$\sqrt{\theta}$ due to the uncertainty.  The
noncommutative geometry is an intrinsic
property of spacetime and does not depend on
any particular feature such as curvature.

To connect the noncommutative geometry to dark
matter and hence to galactic rotation curves,
we need to introduce the metric for a static
spherically symmetric spacetime:
\begin{multline}\label{E:line1}
ds^{2}=-e^{\nu(r)}dt^{2}+e^{\lambda(r)}dr^{2}\\
+r^{2}(d\theta^{2}+\text{sin}^{2}\theta\,d\phi^{2}).
\end{multline}
For this metric, the Einstein field equations are
\begin{equation}\label{E:Einstein1}
e^{-\lambda}
\left[\frac{\lambda^\prime}{r} - \frac{1}{r^2}
\right]+\frac{1}{r^2}= 8\pi \rho,
\end{equation}

\begin{equation}\label{E:Einstein2}
e^{-\lambda}
\left[\frac{1}{r^2}+\frac{\nu^\prime}{r}\right]
-\frac{1}{r^2}= 8\pi p_r,
\end{equation}

\noindent and

\begin{multline}\label{E:Einstein3}
\frac{1}{2} e^{-\lambda} \left[\frac{1}{2}(\nu^\prime)^2+
\nu^{\prime\prime} -\frac{1}{2}\lambda^\prime\nu^\prime +
\frac{1}{r}({\nu^\prime- \lambda^\prime})\right]\\
=8\pi p_t.
\end{multline}

One goal of any modified gravitational theory is
to explain the peculiar behavior of galactic
rotation curves without postulating the existence
of dark matter: test particles move with constant
tangential velocity $v^{\phi}$ in a circular path.
It is noted in Ref. \cite{BHL08} that galactic
rotation curves generally show much more
complicated dynamics.  For present purposes,
however, the analysis can be restricted to the
region in which the velocity is indeed constant.
So taking the observed flat rotation curves as
input, it is well known that, as a result,
\begin{equation}
  e^{\nu}=B_0r^l,
\end{equation}
where $l=2v^{2\phi}$ and $B_0$ is an integration
constant \cite{kN09a}.  Moreover, it is shown
in Ref. \cite{MGL00} that in the presumed dark
matter dominated region, $v^{\phi}\sim
300 \,\text{km/s}=10^{-3}$ for a typical galaxy.
So $l=0.000001$ \cite{kN09b}.  (We are using
units in which $c=G=1.$)

To address the issue of stable orbits, we first
note that given the four-velocity $U^{\alpha}=
dx^{\alpha}/d\tau$ of a test particle moving
solely in the ``equatorial plane" $\theta=\pi /2$
of the galactic halo, the equation
$g_{\nu\sigma}U^{\nu}U^{\sigma}=-m_0^2$ can be
cast in the Newtonian form
\begin{equation}
  \left(\frac{dr}{d\tau}\right)^2=E^2+V(r),
\end{equation}
which results in
\begin{equation}\label{E:potential}
   V(r)=-E^2+E^2\frac{e^{-\lambda}}{B_0r^l}-
   e^{-\lambda}\left(1+\frac{L^2}{r^2}\right).
\end{equation}
Here the constants $E$ and $L$ are, respectively,
the conserved relativistic energy and angular
momentum per unit rest mass of the test particle
\cite{kN09b}.  We are going to define circular
orbits by $r=R_0$, a constant.  We now have
\begin{equation}
    \frac{dR_0}{d\tau}=0 \quad \text{and} \quad
    \left.\frac{dV}{dr}\right|_{r=R_0}=0.
\end{equation}
From these conditions, we obtain \cite{fR12}
\begin{equation}\label{E:L}
   L=\pm\sqrt{\frac{l}{2-l}}R_0
\, \text{and}\,
   E=\pm\sqrt{\frac{2B_0}{2-l}}R_0^{l/2}.
\end{equation}
The orbits are stable if
\begin{equation}\label{E:stable1}
   \left.\frac{d^2V}{dr^2}\right|_{r=R_0}<0
\end{equation}
and unstable if
\begin{equation}\label{E:stable1}
   \left.\frac{d^2V}{dr^2}\right|_{r=R_0}>0.
\end{equation}

\section{The solution}\label{S:solution}
The smeared gravitational source in Eq.
(\ref{E:rho}) leads to a smeared mass.  More
precisely, the Schwarzschild solution of the
Einstein field equations associated with the
smeared source leads to the line element
\begin{multline}\label{E:line2}
ds^{2}=\\
-\left(1-\frac{2M_{\theta}(r)}{r}\right)dt^{2}
  +\left(1-\frac{2M_{\theta}(r)}{r}\right)^{-1}dr^{2}\\
+r^{2}(d\theta^{2}+\text{sin}^{2}\theta\,d\phi^{2}).
\end{multline}
The smeared mass is implicitly given by
\begin{multline}\label{E:Mtheta}
   M_{\theta}(r)=\int^r_0\rho(r')4\pi (r')^2\,dr'
  \\ = \frac{2M}{\pi}\left(\text{tan}^{-1}
  \frac{r}{\sqrt{\theta}}
  -\frac{r\sqrt{\theta}}{r^2+\theta}\right),
\end{multline}
which can also be obtained from Eq.
(\ref{E:Einstein1}).  (Eqs. (\ref{E:Einstein2})
and (\ref{E:Einstein3}) also yield $p_r$ and
$p_t$, as in Ref. \cite{fR12}, but are not
needed for present purposes.)  Due to the
smearing, the mass of the particle depends
on $\theta$, as one would expect.  As in the
case of the Gaussian model, the mass of the
particle is zero at the center and rapidly
rises to $M$.  So from a distance, the smearing
is no longer apparent and we get an ordinary
particle.  In other words,
\begin{equation}
  \text{lim}_{\theta\rightarrow 0}M_{\theta}=M,
\end{equation}
so that the modified Schwarzschild solution
reduces to the ordinary Schwarzschild solution.
(See Fig. 1.)
\begin{figure}[tbp]
\begin{center}
\includegraphics[width=0.5\textwidth]
   {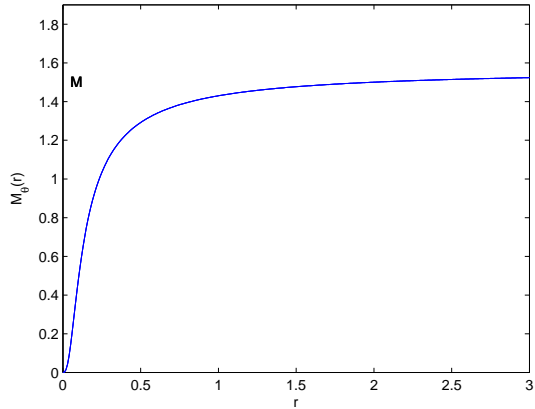}
\end{center}
\caption{The graph of the smeared mass
   $M_{\theta}(r)$.}
\end{figure}

The mass $M$ could be a diffused centralized
object.  Since we are interested in galactic
rotation curves at some fixed distance $r=R_0$
from the center, we will consider instead a
thin spherical shell of radius $r=R_0$.  So
instead of a smeared object, we have a
smeared spherical surface.  We consider the
smearing in the outward radial direction
only, that being the analogue of the smeared
particle at the origin.  It follows that
$\rho(r)$ in Eq. (\ref{E:rho}) must be
replaced by the translated function
\begin{equation}
    \rho(r)=\frac{M\sqrt{\theta}}
    {\pi^2[(r-R_0)^2+\theta]^2}.
\end{equation}

Observe that the mass of the shell becomes
\begin{multline}\label{E:m(r)}
   m(r)=\\ \frac{2M}{\pi}
   \left[\text{tan}^{-1}\frac{r-R_0}
   {\sqrt{\theta}}-\frac{(r-R_0)\sqrt{\theta}}
   {(r-R_0)^2+\theta}\right],
\end{multline}
again dependent on $\theta$ (Fig. 2).  Also
\begin{figure}[tbp]
\begin{center}
\includegraphics[width=0.5\textwidth]{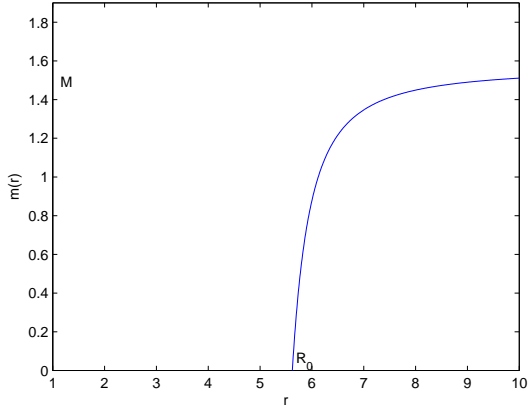}
\end{center}
\caption{The graph of $m(r)$.}
\end{figure}
analogous is $\text{lim}_{\theta\rightarrow 0}
\,m(r)=M$, where $M$ is now the mass of the shell.
So in geometrized units, $M$ and $m(r)$ are much
less than $R_0$.

 At this point we can finally address the
 question of stability by examining the
 potential $V(r)$ more closely.  In view of line
 element (\ref{E:line2}), we now have
 \begin{equation}
      e^{-\lambda}=1-\frac{2m(r)}{r}.
 \end{equation}
So from Eq. (\ref{E:potential}),
\begin{multline}
V(r)=-E^2+E^2\frac{r^{-l}}{B_0}\left(1-\frac{2m}{r}
  \right)\\ -\left(1-\frac{2m}{r}\right)
      \left(1+\frac{L^2}{r^2}\right).
\end{multline}
To see the effect of the smearing, we first compute
$V''(r)$:
\begin{multline}\label{E:Vup}
V''(r)=
  \frac{2R_0^2\,l(l+1)}{2-l}\frac{1}{r^{l+2}}
  \left(1-\frac{2m}{r}\right)\\-\frac{6lR_0^2}{2-l}
  \frac{1}{r^4}\left(1-\frac{2m}{r}\right)\\+
  \frac{8lR_0^2}{2-l}\frac{1}{r^{l+1}}
  \frac{rm'-m}{r^2}\\-\frac{8lR_0^2}{2-l}
  \frac{1}{r^3}\frac{rm'-m}{r^2}\\-\frac{4R_0^l}{2-l}
  \frac{1}{r^l}\frac{r^2m''-2rm'+2m}{r^3}\\+
  2\left(1+\frac{lR_0^2}{2-l}\frac{1}{r^2}\right)
  \frac{r^2m''-2rm'+2m}{r^3}.
  \end{multline}
From Eq. (\ref{E:m(r)}),
\begin{equation}\label{E:mprime}
    m'(r)=\frac{4M}{\pi}\frac{(r-R_0)^2\sqrt{\theta}}
    {[(r-R_0)^2+\theta]^2}
\end{equation}
and
\begin{equation}\label{E:mdoubleprime}
   m''(r)=\frac{8M}{\pi}
   \frac{(r-R_0)\sqrt{\theta}-(r-R_0)^3\sqrt{\theta}}
   {[(r-R_0)^2+\theta]^3}.
\end{equation}
It now follows directly that at $r=R_0$, only
the first two terms in Eq. (\ref{E:Vup}) are
nonzero:
\begin{multline}\label{E:atRzero}
   V''(R_0)=\frac{2R_0^l\,l(l+1)}{2-l}\frac{1}
   {R_0^{l+2}}-\frac{6lR_0^2}{2-l}\frac{1}{R_0^4}
   \\ \approx -\frac{4l}{(2-l)R_0^2}<0.
\end{multline}
We therefore have a stable orbit at $r=R_0$ due to
the noncommutative geometry.

\section{The need for noncommutative geometry}
  \label{S:necessary}
We saw in the previous section that the smearing
effect in noncommutative geometry is responsible
for the stable orbit at $r=R_0$.  In this section
we study the effect of reduced smearing (due to
diminishing $\theta$), thereby approaching
Einstein gravity.  The idea is to show that in
this limit, the stability criterion is no longer
met.

To this end, we return to Eq. (\ref{E:Vup})
and observe that the third term,
\begin{equation}\label{E:z1}
   z(r,\theta)=\frac{8lR_0^2}{2-l}\frac{1}
   {r^{l+1}}\frac{rm'-m}{r^2}
\end{equation}
strongly dominates near $r=R_0$ since the
denominator is much smaller than the denominator
in all the other terms.  As we saw, at $r=R_0$,
both $m'$ and $m$ are equal to zero, but $rm'-m$
is positive for $r>R_0$ and, as we will see
later in Fig. 3, sharply increasing near
$r=R_0$ for any fixed $\theta$.  So the positive
third term easily catches up with the (negative)
sum of the first two terms.  Given that the
remaining terms are negligible, we can now say
that there exists an $r=r_1$ (for every
$\theta$) such tht
\begin{equation}\label{E:z2}
   \left. z(r,\theta)-\frac{4l}{(2-l)R_0^2}
   \right|_{r=r_1}=0.
\end{equation}
Hence $V''(r_1)=0$ and (for every $\theta$),
$V''(r)<0$ in the interval $[R_0,r_1]$
and $V''(r)>0$ for $r>r_1$.  These cases
will be discussed separately.

\subsection{$V''(r)<0$}
As noted above, for $r<r_1$, we have $V''(r)<0$,
where $\theta$ is assumed fixed.  We wish to
show that an ever smaller $\theta$ results in
an ever smalller interval $[R_0,r]$ for which
$V''(r)<0$.

To this end, we obtain from Eq. (\ref{E:z2}),
\begin{equation}\label{E:plane1}
   \left. \frac{8lR_0^2}{(2-l)r^{l+1}}
   \frac{rm'-m}{r^2}-\frac{4l}{(2-l)R_0^2}
   \right|_{r=r_1}=0
\end{equation}
and hence from Eqs. (\ref{E:mprime}) and
(\ref{E:mdoubleprime}),
\begin{multline}\label{E:plane2}
    \frac{2M}{\pi}\frac{8lR_0^2}{(2-l)
   r^{l+1}}\left[\frac{2r(r-R_0)^2\sqrt{\theta}}
   {[(r-R_0)^2+\theta]^2}\right.\\ \left.-\text{tan}^{-1}
   \frac{r-R_0}{\sqrt{\theta}}+\frac{(r-R_0)
   \sqrt{\theta}}{(r-R_0)^2+\theta}\right]\\
   \left. -\frac{4l}{(2-l)R_0^2}\right|_{r=r_1}
   =0
\end{multline}
for every fixed $\theta$.  So to study the
relationship between $\theta$ and $r$ qualitatively,
we can choose an arbitrary ray and consider the
first term in Eq. (\ref{E:plane2}),
\begin{multline}\label{E:Q}
   Q(r,\theta)= \frac{2M}{\pi}\frac{8lR_0^2}{(2-l)
   r^{l+1}}\times\\ \left[\frac{2r(r-R_0)^2\sqrt{\theta}}
   {[(r-R_0)^2+\theta]^2}\right.\\ -\left.\text{tan}^{-1}
   \frac{r-R_0}{\sqrt{\theta}}+
   \frac{(r-R_0)\sqrt{\theta}}{(r-R_0)^2+\theta}
   \right ]
\end{multline}
as a function of $r$ and $\theta$ in rectangular
coordinates.  The condition in Eq. (\ref{E:plane2})
can now be viewed as the plane $Q=4l/(2-l)R_0^2$
passing through the surface $Q=Q(r,\theta)$.  The
resulting relationship between $\theta$ and $r$
in this plane is not a simple one-to-one
correspondence because the intersection is
oval-shaped.  However, we know that for physical
reasons, $\theta$ is necessarily small and $r$
close to $r=R_0$.

So, as a next step, we plot $Q(r,\theta)$ in
Eq. (\ref{E:Q}) for a few values of $\theta$,
intersected by the line $Q=4l/(2-l)R_0^2$,
shown in Fig. 3.  For each curve, the
\begin{figure}[tbp]
\begin{center}
\includegraphics[width=0.5\textwidth]{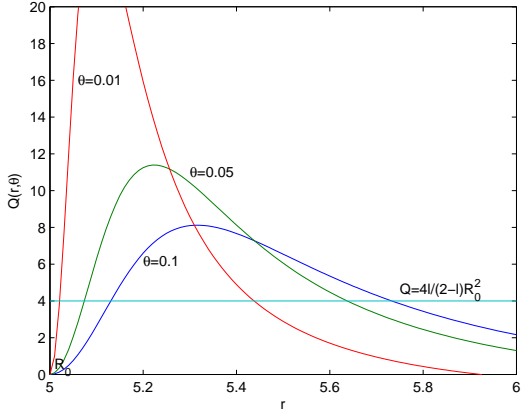}
\end{center}
\caption{Plot showing that if $\theta\rightarrow 0$,
    then $r_1\rightarrow R_0$.}
\end{figure}
intersection is at $r=r_1$.  As already noted, there
are indeed two values of $r$ for every fixed
$\theta$, but only the smaller value is physically
relevant.  Fig. 3  shows that if $\theta\rightarrow
0$, then the left side of Eq. (\ref{E:Q}) can remain
fixed only if $r_1\rightarrow R_0$.  By continuity,
then, $V''(r)\ge 0$ for $r\ge R_0$, i.e., the
stability criterion is no longer satisfied.  We
conclude that noncommutative geometry is not only
sufficient but also necessary for meeting the
stability criterion.  Without the
noncommutative-geometry background, the stability
of the orbit would have to be attributed to
another cause, such as dark matter.

\subsection{$V''(r)>0$}
Recall that $V''(r)>0$ for $r>r_1$, for any
fixed $\theta$.  So outside the smeared region,
the stability criterion is no longer met, even
though we have a stable orbit at $r=R_0$.  The
implication is  that from a distance, the smearing is
no longer apparent, even though it is still
very much present.  So, in a sense, the unseen
dark matter is replaced by the unseen
noncommutative geometry.

\subsection{Conclusion}
It is shown in Ref. \cite{fR12} that a
noncommutative-geometry background can account
for galactic rotation curves without the need for
dark matter.  The smearing effect that
characterizes noncommutative geometry is described
by means of a Gaussian distribution of minimal
length $\sqrt{\theta}$.  The purpose of this paper
is two-fold: (1) to confirm the conclusions in
Ref. \cite{fR12} in a simpler and more intuitive
way by using the distribution proposed in Ref.
\cite{NM08} instead of the Gaussian function,
and (2) to show that the smearing effect is
both a necessary and sufficient condition for
meeting the stability criterion.

That noncommutative geometry, which has all
the appearances of a small effect, can account
for the galactic rotation curves is consistent
with the corresponding situation in $f(R)$
gravity: only a small change in the Ricci
scalar is required to account for dark
matter \cite{BHL08}.

\end{document}